\documentclass{mn2e}
\usepackage{epsfig}

\title[Carina's Defiant Finger]{Carina's Defiant Finger: {\it HST}
Observations of a Photoevaporating Globule in NGC~3372\thanks{Based in
part on observations made with the NASA/ESA {\it Hubble Space
Telescope}, obtained at the Space Telescope Science Institute, which
is operated by the Association of Universities for Research in
Astronomy, Inc., under NASA contract NAS5-26555.}}

\author[N.\ Smith et al.]{Nathan Smith$^1$\thanks{Hubble Fellow;
nathans@casa.colorado.edu}\thanks{Visiting Astronomer, Cerro Tololo
Inter-American Observatory, National Optical Astronomy Observatory,
operated by the Association of Universities for Research in Astronomy,
Inc., under cooperative agreement with the National Science
Foundation.}, Rodolfo H.\ Barb\'{a}$^2$, and Nolan R.\ Walborn$^3$ \\
$^1$Center for Astrophysics and Space Astronomy, University of
Colorado, 389 UCB, Boulder, CO 80309, USA \\ $^2$Departmento de
F\'{i}sica, Universidad de La Serena, Benavente 980, La Serena, Chile
\\ $^3$Space Telescope Science Institute, 3700 San Martin Drive,
Baltimore, MD 21218, USA}

\date{Accepted 0000, Received 0000, in original form 0000}
\pagerange{\pageref{firstpage}--\pageref{lastpage}} \pubyear{2002}

\def\arcdeg{\degr}

\begin{document}
\label{firstpage}
\maketitle
\begin{abstract}

We present {\it Hubble Space Telescope} Wide Field Planetary Camera 2
images of a prominent externally-ionized molecular globule in the
Carina Nebula (NGC~3372), supplemented with ground-based infrared
images and visual-wavelength spectra.  This molecular globule has a
shape resembling a human hand, with an extended finger that points
toward its likely source of ionizing radiation.  Following an analysis
of the spatially-resolved ionization structure and spectrum of the
photoevaporative flow from the Finger, we conclude that the dominant
ionizing source is either the WNL star WR25 (HD~93162), the adjacent
O4~If-type star Tr16-244, or perhaps both.  We estimate a mass-loss
rate of $\sim$2$\times$10$^{-5}$ M$_{\odot}$ yr$^{-1}$ from the main
evaporating surface of the globule, suggesting a remaining lifetime of
10$^{5.3}$ to 10$^6$ years. We find a total mass for the entire
globule of more than 6 M$_{\odot}$, in agreement with previous
estimates. The hydrogen column density through the globule derived
from extinction measurements is a few times 10$^{22}$ cm$^{-2}$, so
the photodissociation region behind the ionization front should be
limited to a thin layer compared to the size of the globule, in
agreement with the morphology seen in H$_2$ images.  Although a few
reddened stars are seen within the boundary of the globule in
near-infrared continuum images, these may be background stars.  We do
not detect a reddened star at the apex of the finger, for example,
down to a limiting magnitude of $m_K\simeq$17.  However, considering
the physical properties of the globule and the advancing ionization
front, it appears that future star formation is likely in the Finger
globule, induced by radiation-driven implosion.

\end{abstract}

\begin{keywords} 
H~{\sc ii} regions --- ISM: globules --- ISM: individual (NGC~3372)
--- stars: formation
\end{keywords}

\section{INTRODUCTION}

Compact bright-rimmed molecular globules are common in evolved massive
star forming regions, and are among the last vestiges of the molecular
cloud that gave birth to the OB stars powering the H~{\sc ii} region
(e.g., Bok \& Reilly 1947).  These globules are externally illuminated
and are photoevaporated and photoionized by UV radiation.  They
typically range in size from 0.1 to 1 pc, and appear striking in
[S~{\sc ii}] $\lambda\lambda$6717,6731 and H$\alpha$ emission arising
in the limb-brightened surfaces of their thin ionization fronts.
Famous examples of this class of objects are Thackeray's globules in
IC~2944 (Thackeray 1950; Reipurth et al.\ 2003), as well as similar
features in the Rosette Nebula (Herbig 1974), the Gum Nebula (Hawarden
\& Brand 1976; Reipurth 1983), and the Carina Nebula (Walborn 1975;
Cox \& Bronfman 1995).

\begin{figure*}\begin{center}
\epsfig{file=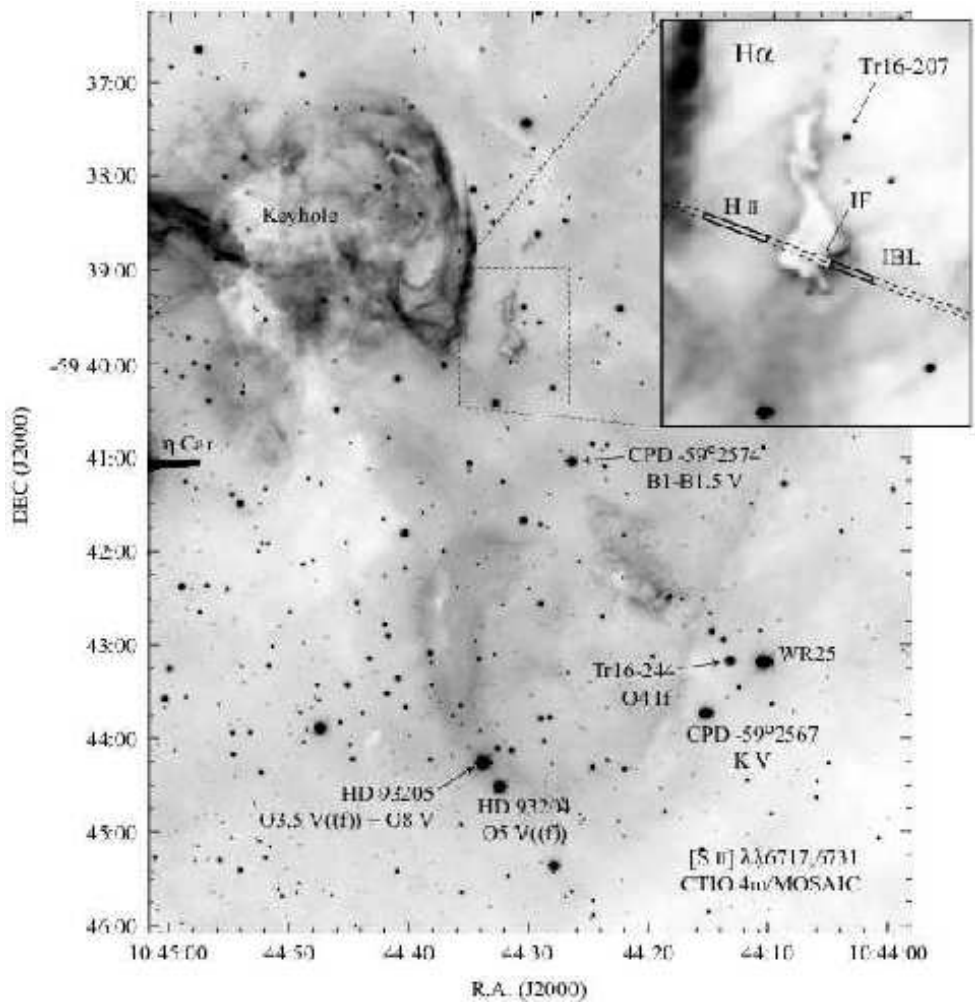,width=5.1in}\end{center}
\caption{Large-scale image of the environment around the Finger,
including the Keyhole and the background Carina nebula, $\eta$ Car,
WR25, and other stars.  The [S~{\sc ii}] image shown here was obtained
with the MOSAIC camera on the CTIO 4m telescope.  The inset shows an
enlargement of the Finger in the H$\alpha$ filter, along with the
orientation of the slit apertures used for spectra in Figure 5.}
\end{figure*}

Material evaporated from these clumps partly fills the interior of an
H~{\sc ii} region and may help stall the advance of the main
ionization front by absorbing incident Lyman continuum radiation and
converting it to a recombination front, but perhaps the most pertinent
role played by these globules is that they are potential sites of
continued star formation.  Some globules show clear evidence for
embedded star formation, while others do not (Reipurth 1983; Reipurth
et al.\ 2003).  If they are sites of star formation, one possibility
is that they are simply dense cores that spontaneously collapsed to
form stars and were then uncovered by the advancing ionization front.
Another more intriguing idea is that pressure from the
ionization-shock front at the surface propagates through a globule and
helps to overcome the magnetic, turbulent, and thermal pressure that
supports it against collapse, thereby inducing a star to form in the
globule.  This process of radiation-driven implosion is often referred
to as {\it triggered star formation}.  However, direct and unambiguous
observational evidence for actively triggered star formation remains
elusive; the pillars in M16 may be an example of this phenomenon
(White et al.\ 1999; Hester et al.\ 1996; Williams et al.\ 2001;
McCaughrean \& Andersen 2002).  Numerous investigators have examined
theoretical aspects of these evaporating globules, including details
of the photoevaporative flows, their effect on the surrounding H~{\sc
ii} region, the globule's shaping, destruction, and acceleration
(through the ``rocket effect''), and the possibility of triggered
collapse and star formation due to the pressure of the impinging
ionization front (Oort \& Spitzer 1955, Kahn 1969; Dyson 1973; Dyson
et al.\ 1995; Elmegreen 1976; Bertoldi 1989; Bertoldi \& McKee 1990;
Bertoldi \& Draine 1996; Lizano et al.\ 1996; Gorti \& Hollenbach
2002; Williams 1999; Williams et al.\ 2001).  The evolution of these
globules depends strongly on the incident UV radiation field, the
initial size and density of the globule, and other properties, so
observations of them in a wide variety of different environments are
useful.

\begin{table*}\begin{minipage}{6in}
\caption{Observations of the Finger}
\begin{tabular}{@{}lllcl}\hline\hline
Telescope &Instrument &Filter or     &Exp.\ Time &Comment \\
          &           &Emiss.\ Lines &(sec)      &        \\ \hline
{\it HST} &WFPC2   &F502N, [O~{\sc iii}] $\lambda$5007            &320  &image \\
{\it HST} &WFPC2   &F656N, H$\alpha$, [N~{\sc ii}]                &400  &image \\
{\it HST} &WFPC2   &F673N, [S~{\sc ii}] $\lambda\lambda$6717,6731 &800  &image \\
CTIO 4m   &OSIRIS  &J                                      &120  &image \\
CTIO 4m   &OSIRIS  &H                                      &120  &image \\
CTIO 4m   &OSIRIS  &K                                      &120  &image \\
CTIO 4m   &OSIRIS  &He~{\sc i} $\lambda$10830              &720  &image \\
CTIO 4m   &OSIRIS  &Pa $\beta$                             &720  &image \\
CTIO 4m   &OSIRIS  &H$_2$ 1-0 S(1) 2.122 $\micron$         &1080 &image \\
CTIO 1.5m &RC Spec &blue; 3600-7100 \AA &1200  &long-slit, P.A.=69$\arcdeg$ \\
CTIO 1.5m &RC Spec &red;  6250-9700 \AA &1200  &long-slit, P.A.=69$\arcdeg$ \\
\hline
\end{tabular}\end{minipage}
\end{table*}

A rich population of these globules resides in the Carina Nebula
(NGC~3372; d=2.3 kpc), with the most notable grouping clustered around
the famous Keyhole Nebula.  Their emission from ionized,
photodissociated, and molecular gas has been documented by several
studies (Walborn 1975; Deharveng \& Maucherat 1975; Cox \& Bronfman
1995; Brooks et al.\ 2000; Smith 2002; Rathborne et al.\ 2002).  There
are a dozen of these clumps around the Keyhole, with typical masses of
order 10 M$_{\sun}$ (Cox \& Bronfman 1995).  A few of these are seen
only in silhouette, but most have bright H$_2$ and polycyclic aromatic
hydrocarbon (PAH) emission from their illuminated surfaces (Brooks et
al.\ 2000; Rathborne et al.\ 2002).  In addition, several dozen
smaller features with diameters of only $\sim$5000 AU were recently
discovered by Smith et al.\ (2003a), and additional proplyd candidates
are seen in our images (see below, and Barb\'{a} et al.\ in prep.).
They may be either large photoevaporating circumstellar disks
(proplyds) or very small cometary clouds that are remnants of larger
globules in more advanced stages of photoevaporation (e.g., Bertoldi
\& McKee 1990).  It is not yet known if these smaller proplyd
candidates or the larger globules in Carina contain embedded
protostars or collapsing cores that may someday form stars.  However,
one large globule in the southern part of the nebula has recently been
shown to harbor a Class~{\sc i} protostar that drives the HH~666 jet
(Smith et al.\ 2004).  In any case, these globules present a rare
opportunity to study their associated phenomena in an environment
powered by some of the hottest and most massive stars in the Galaxy
(Walborn 1995; Walborn et al.\ 2002), while still being located near
to us and suffering little interstellar extinction.

In this paper we untertake a detailed analysis of one particularly
striking bright-rimmed globule in the Carina Nebula (see Figure 1)
that we call the ``Finger'' because of its gesticulatory morphology
(it has previously been identified as clump 4; Cox \& Bronfman 1995).
The defiant finger seems to point toward its dominant UV source, and
as such, allows for an interesting application of models of
photoevaporating globules mentioned above.  In \S 2 we discuss our
observations, and in \S 3 and \S 4 we briefly discuss the results of
our imaging and spectroscopy.  In \S 5 we undertake a detailed
analysis of the photoevaporative flow, and in \S 6 we discuss
implications for potential star formation in the globule.

\begin{figure*}\begin{center}
\epsfig{file=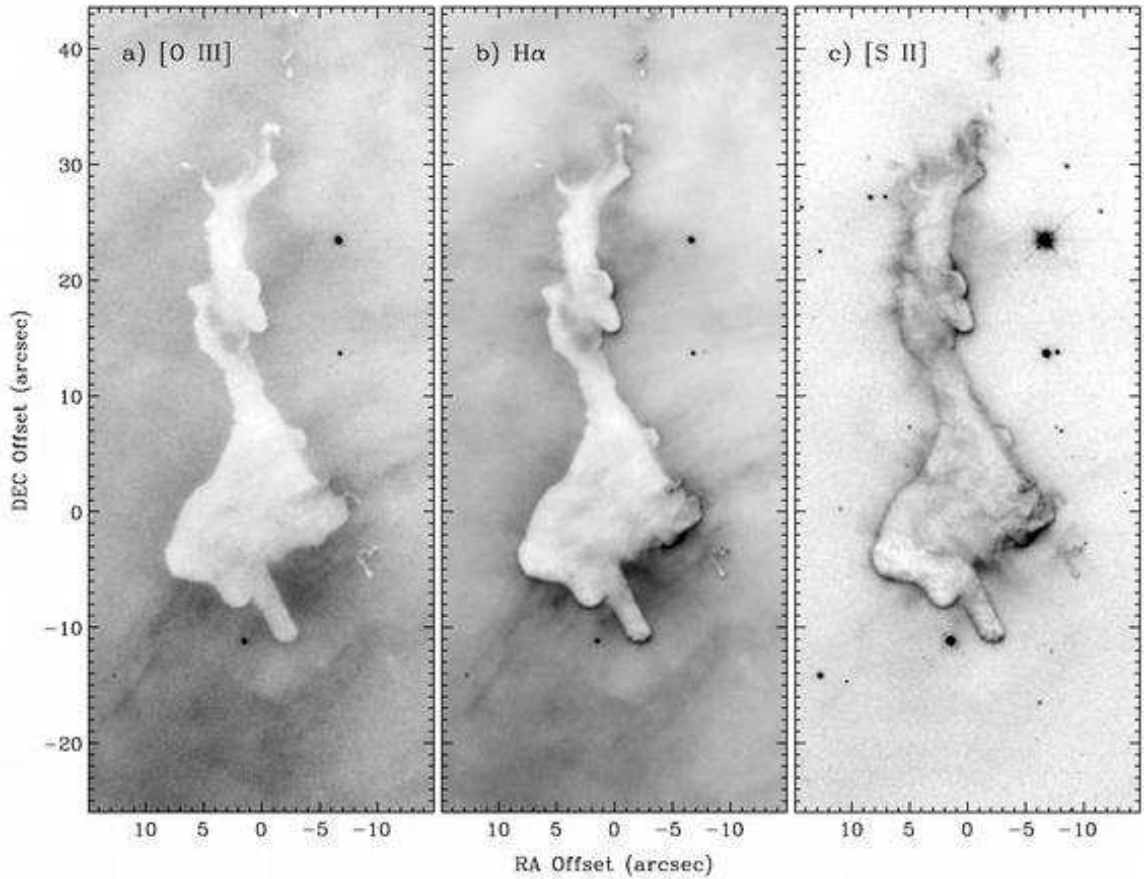,width=6in}\end{center}
\caption{{\it HST}/WFPC2 images of the Finger.  (a) F502N filter
transmitting [O~{\sc iii}] $\lambda$5007; grayscale range from
2.4$\times$10$^{-14}$ (white) to 1.9$\times$10$^{-13}$ (black) ergs
s$^{-1}$ cm$^{-2}$ arcsec$^{-2}$. (b) F656N filter transmitting
H$\alpha$; grayscale range from 5.4$\times$10$^{-14}$ (white) to
3$\times$10$^{-13}$ (black) ergs s$^{-1}$ cm$^{-2}$ arcsec$^{-2}$. (c)
F673N filter transmitting [S~{\sc ii}] $\lambda\lambda$6717,6731;
grayscale range from 4.2$\times$10$^{-15}$ (white) to
3.3$\times$10$^{-14}$ (black) ergs s$^{-1}$ cm$^{-2}$ arcsec$^{-2}$.
These observed flux levels were not corrected for reddening and
extinction.  The axes display offset in arcsec from an arbitrary
central position in the globule.}
\end{figure*}

\begin{figure*}\begin{center}
\epsfig{file=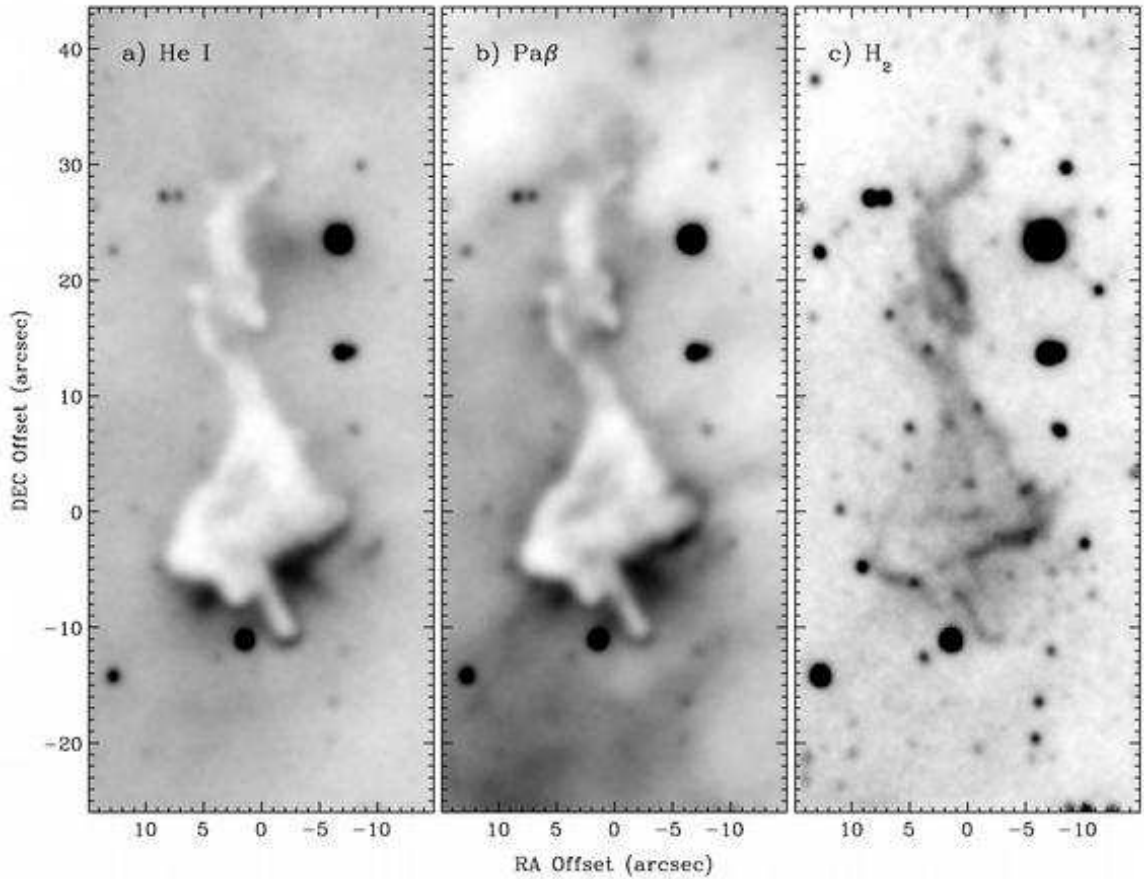,width=6in}\end{center}
\caption{Near-IR images of the Finger.  (a) He~{\sc i} $\lambda$10830;
grayscale range from 1.3$\times$10$^{-14}$ (white) to
4.8$\times$10$^{-14}$ (black) ergs s$^{-1}$ cm$^{-2}$
arcsec$^{-2}$. (b) Hydrogen Pa$\beta$; grayscale range from
1.5$\times$10$^{-14}$ (white) to 5.1$\times$10$^{-14}$ (black) ergs
s$^{-1}$ cm$^{-2}$ arcsec$^{-2}$. (c) H$_2$ $v=1-0$ S(1)
$\lambda$21218; grayscale range from 9.3$\times$10$^{-16}$ (white) to
5.3$\times$10$^{-15}$ (black) ergs s$^{-1}$ cm$^{-2}$ arcsec$^{-2}$.
These observed flux levels were not corrected for reddening and
extinction.}
\end{figure*}

\begin{figure*}\begin{center}
\epsfig{file=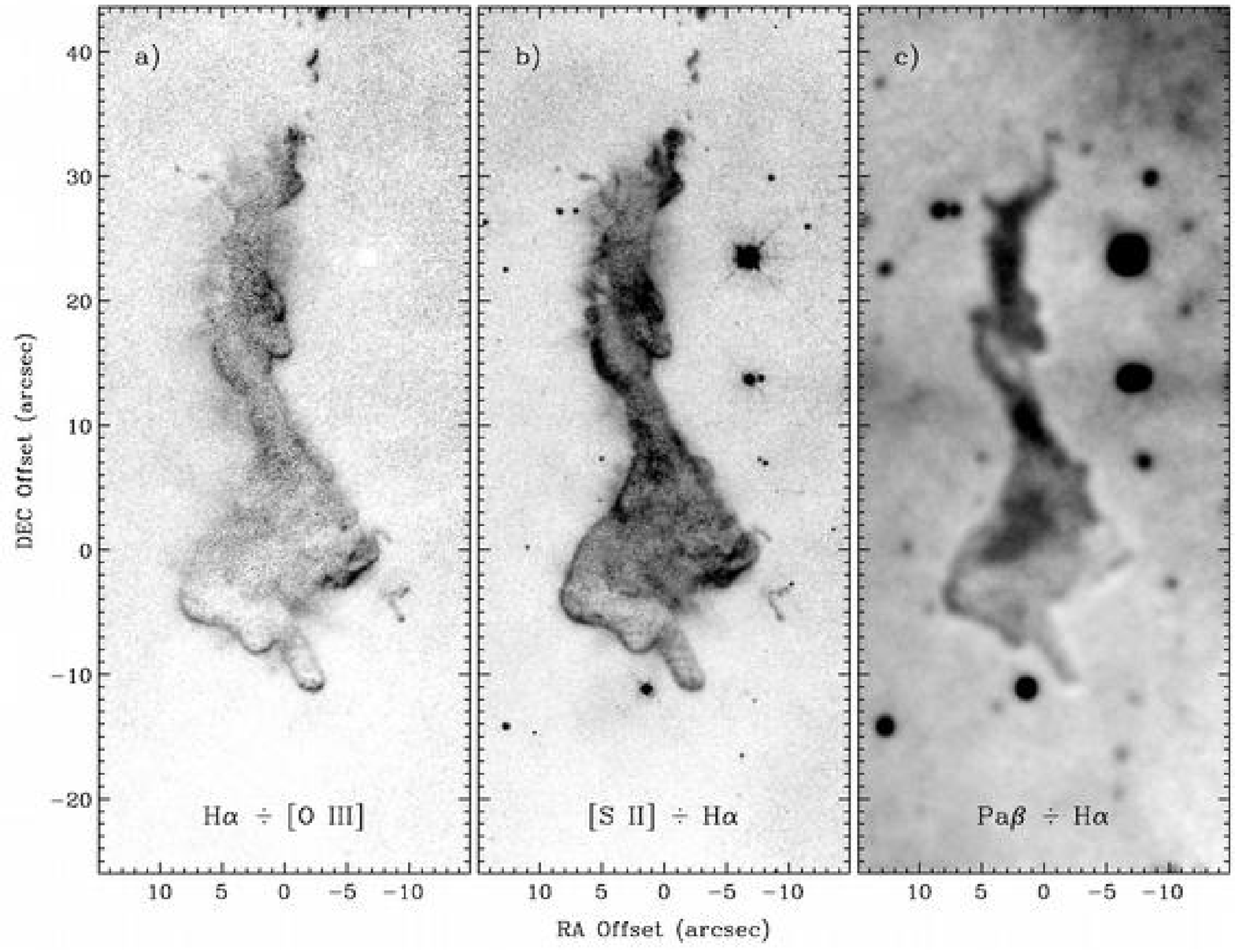,width=6in}\end{center}
\caption{Selected flux-ratio images of the Finger.  (a)
H$\alpha\div$[O~{\sc iii}] (F656N$\div$F502N) with the grayscale range
from 1 (white) to 2.3 (black).  (b) [S~{\sc ii}]$\div$H$\alpha$
(F673N$\div$F656N) with the grayscale range from 0.033 to 0.13.  (c)
The ground-based Pa$\beta$ image divided by a smoothed {\it HST}/WFPC2
H$\alpha$ image, with the grayscale range from 0.07 to 0.12.  Note
that the F656N H$\alpha$ image is partially contaminated by adjacent
[N~{\sc ii}] lines, which affects the value of the flux ratios listed
here.  These flux ratios have been corrected for reddening and
extinction, using $E(B-V)$=0.37 and $R$=4.8 (see text).}
\end{figure*}

\section{OBSERVATIONS AND DATA REDUCTION}

\subsection{{\it HST}/WFPC2 Images}

Images of the Keyhole Nebula were obtained on 1999 April 18 with the
Wide Field Planetary Camera 2 (WFPC2) instrument onboard the {\it
Hubble Space Telescope} ({\it HST}).  Four different pointings were
observed and combined to form a large mosaic image, which has been
released by the Hubble Heritage Team\footnotemark\footnotetext{See
{\tt http://oposite.stsci.edu/pubinfo/pr/2000/06/}.}.  Here we focus
on only a small portion of the total area observed, containing the
globule that we call the Finger.  Images were obtained through
narrowband filters F502N ([O~{\sc iii}] $\lambda$5007), F656N
(H$\alpha$), and F673N ([S~{\sc ii}] $\lambda\lambda$6717,6731), as
well as the broadband filters F469W, F555W, and F814W.  Only the
emission-line images will be discussed here, however.  Pairs of images
in each narrowband filter were combined and processed in the usual way
to remove cosmic rays and hot pixels, and to correct for geometric
distortion, as described in the {\it HST} Data Handbook (however, the
images were not CTE corrected).  Finally, the individual frames were
combined to form a single mosaic image with a pixel scale of
0$\farcs$0995.  Subsections of the images showing the region around
the Finger are displayed in Figure 2, oriented with north up and east
to the left.

\subsection{Near-Infrared Images}

Near-infrared (IR) images of the Finger and its immediate environment
were obtained on 2003 February 25 with the Ohio State IR Imaging
Spectrometer (OSIRIS) mounted on the CTIO 4m telescope.  OSIRIS has a
1024$\times$1024 NICMOS3 array, with a pixel scale of 0$\farcs$161.
Only a portion of the array is illuminated in this configuration,
yielding a field of view of $\sim$1$\farcm$5.  We obtained images in
the broadband $J$, $H$, and $K$ filters, as well as narrowband filters
isolating He~{\sc i} $\lambda$10830, Pa$\beta$ $\lambda$12818, and
H$_2$ 1-0 S(1) $\lambda$21218.  In each filter, several individual
exposures were taken with small offsets between them to correct for
bad pixels on the array.  The seeing during the observations was
$\sim$0$\farcs$8.  There is no position on the sky near the Finger
that is suitable for measuring the background sky because of bright
emission from the H~{\sc ii} region and nearby star clusters, so for
sky subtraction we nodded the telescope to a position roughly
2$\arcdeg$ south.  Flux calibration was accomplished by measuring
isolated stars from the 2MASS catalog that were included in our field
of view, and is valid to within about $\pm$10\%.  The IR images were
aligned with the WFPC2 images and are shown in Figure 3.

\subsection{Optical Spectroscopy}

Low-resolution ($R \ \sim \ 700-1600$) spectra from 3600 to 9700 \AA \
were obtained on 2002 March 1 and 2 using the RC Spectrograph on the
CTIO 1.5-m telescope.  Long-slit spectra of the Finger were obtained
with the 1$\farcs$5-wide slit aperture oriented at
P.A.$\approx$69$\arcdeg$ crossing through the middle of the globule as
shown in Figure 1.  The pixel scale in the spatial direction was
1$\farcs$3.  Spectra were obtained on two separate nights in two
different wavelength ranges (blue, 3600-7100 \AA; and red, 6250-9700
\AA), with total exposure times and other details listed in Table 1.
Sky conditions were mostly photometric, although a few thin transient
clouds were present when the blue spectrum was obtained. Flux
calibration and telluric absorption correction were accomplished using
the standard stars LTT-3218 and LTT-2415.

From the resulting long-slit spectra, we extracted several
one-dimensional (1-D) segments of the slit that sampled emission from
the thin ionization front (IF), the ionized boundary layer (IBL), the
total emission from the IF and IBL (TOT), and a nearby sample of the
emission from the background Carina Nebula H~{\sc ii} region.  The
sizes of these subapertures are identified in Figures 1 and 5.  The
blue and red wavelength ranges of these extracted 1-D spectra were
merged to form a single 3600-9700 \AA \ spectrum for each position,
with a common dispersion of 2 \AA \ pixel$^{-1}$; the average of the
two was taken in the region of the spectrum near H$\alpha$ where the
blue and red spectra overlapped (in the overlap region, bright line
fluxes agreed to within about $\pm$5\%, comparable to the assumed
uncertainty).  Finally, the background H~{\sc ii} region spectrum was
subtracted from the IF, IBL, and TOT spectra.  The final
flux-calibrated spectra are shown in Figure 5.  Observed intensities
are listed in Table 2, relative to H$\beta$=100.  Uncertainties in
these line intensities vary depending on the strength of the line and
the measurement method.  The integrated fluxes of isolated emission
lines were measured; for these, brighter lines with $I>$10 typically
have measurement errors of a few percent, and weaker lines may have
uncertainties of $\pm$10 to 15\%.  The uncertainties increase somewhat
at the blue edge of the spectrum.  Blended pairs or groups of lines
were measured by fitting Gaussian profiles.  For brighter blended
lines like H$\alpha$+[N~{\sc ii}] and [S~{\sc ii}], the measurement
uncertainty is typically 5 to 10\%.  Obviously, errors will be on the
high end for faint lines adjacent to bright lines, and errors will be
on the low end for the brightest lines in a pair or group, or lines in
a pair with comparable intensity.

\section{IMAGING RESULTS}

Figures 2, 3, and 4 show imaging data concentrated on a small region
(less than 1 square arcminute) featuring the evaporating molecular
globule that is the focus of this paper.  Images showing the Keyhole
Nebula and its environment on a much larger spatial scale but with
lower spatial resolution, and using filters transmitting these same
optical and IR emission lines, have been presented by Smith (2002),
Brooks et al.\ (2000), Lopez \& Meaburn (1984), Walborn (1975), and
Deharveng \& Maucherat (1975).  Near-IR continuum images of the Finger
will be discussed later in \S 6.

\begin{figure*}\begin{center}
\epsfig{file=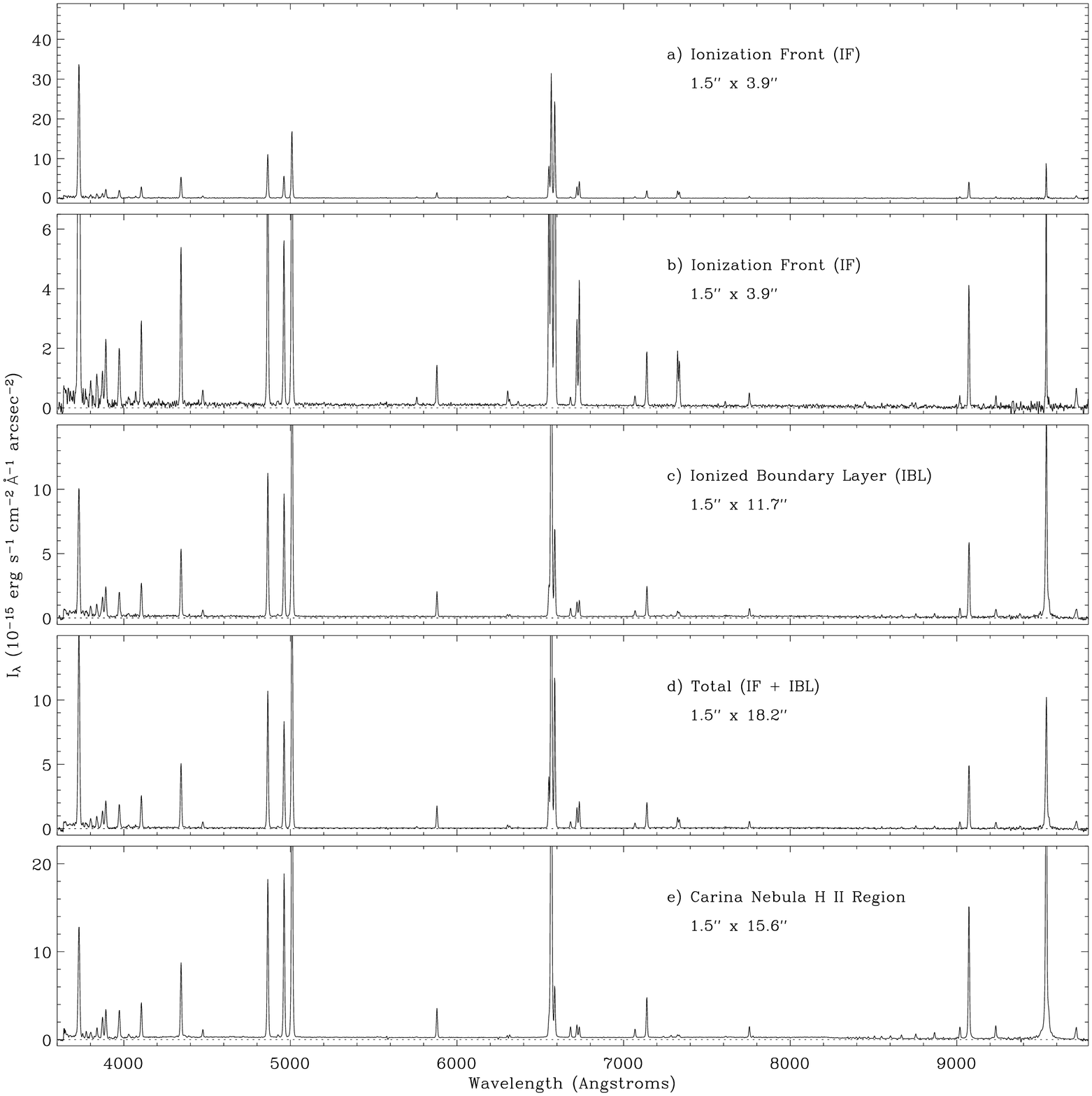,width=5.5in}\end{center}
\caption{Spectra of gas associated with the Finger.  The plotted
quantity is the average specific intensity in the aperture size given
in each panel, and shown in Figure 1.  (a) The spectrum of the
ionization front (IF) on the side of the globule facing the main UV
source.  (b) Same as Panel $a$ but with a different intensity scale.
(c) Spectrum of the ionized boundary layer (IBL, or ``photoevaporative
flow'') associated with the main ionization front.  (d) Spectrum of
the total emission including both the ionization front and the ionized
boundary layer (note that the ``total'' emission shown here is not a
simple average of the IF and IBL; the aperture used was slightly
larger than both the IF and IBL combined; see Figure 1).  (e) Average
spectrum of the background H~{\sc ii} region near the Finger.}
\end{figure*}

\begin{table*}\begin{minipage}{5.5in}
\caption{Observed and Dereddened Line Intensities$^a$}
\begin{tabular}{@{}lcccccccc}\hline\hline
$\lambda$(\AA) &IF    &IF       &IBL   &IBL      &TOT   &TOT      &H~{\sc ii} &H~{\sc ii} \\
and I.D.       &(Obs) &(Dered.) &(Obs) &(Dered.) &(Obs) &(Dered.) &(Obs)      &(Dered.)   \\ \hline
3727 [O~{\sc ii}]	&312	      &436	      &105	      &146	      &178	      &248	      &93.0	      &130      \\
3869 [Ne~{\sc iii}]	&9.9	      &13.3	      &16.2	      &21.9	      &14.0	      &18.8	      &15.0	      &20.1     \\
3889 H8+He~{\sc i}	&19.1	      &25.5	      &22.2	      &29.7	      &21.7	      &29.0	      &20.2	      &27.1     \\
3970 H$\epsilon$+[Ne~{\sc iii}]	&18.1 &23.7	      &19.2	      &25.2	      &20.0	      &26.2	      &20.6	      &27.1     \\
4026 He~{\sc i}		&2.9	      &3.7	      &2.4	      &3.1	      &2.4	      &3.1	      &2.6	      &3.3      \\
4073 [S~{\sc ii}]	&3.9	      &5.0	      &...	      &...	      &1.6	      &2.0	      &0.7	      &1.0      \\
4102 H$\delta$		&24.8	      &31.3	      &23.4	      &29.6	      &23.8	      &30.0	      &23.4	      &29.5     \\
4340 H$\gamma$		&44.6	      &52.6	      &46.3	      &54.7	      &47.5	      &56.1	      &44.9	      &53.0     \\
4363 [O~{\sc iii}]	&0.8	      &0.9	      &1.3	      &1.6	      &0.8	      &0.9	      &4.9	      &5.7      \\
4387 He~{\sc i}		&...          &...	      &0.9	      &1.1	      &0.8	      &0.9	      &0.5	      &0.6      \\
4471 He~{\sc i}		&4.1	      &4.7	      &4.3	      &4.9	      &4.5	      &5.1	      &4.9	      &5.6      \\
4861 H$\beta$		&100	      &100	      &100	      &100	      &100	      &100	      &100	      &100      \\
4922 He~{\sc i}		&1.6	      &1.6	      &1.5	      &1.5	      &1.3	      &1.3	      &1.6	      &1.5      \\
4959 [O~{\sc iii}]	&51.7	      &50.1	      &83.8	      &81.2	      &77.1	      &74.7	      &99	      &96.1     \\
5007 [O~{\sc iii}]	&161	      &153	      &254	      &242	      &236	      &225	      &300	      &286      \\
5755 [N~{\sc ii}]	&3.22	      &2.5	      &0.8	      &0.6	      &1.5	      &1.2	      &0.7	      &0.56     \\
5876 He~{\sc i}		&15.0	      &11.2	      &16.3	      &12.2	      &15.8	      &11.8	      &16.8	      &12.5     \\
6300 [O~{\sc i}]	&5.1	      &3.4	      &1.3	      &0.9	      &2.2	      &1.5	      &0.9	      &0.6      \\
6312 [S~{\sc iii}]	&2.5	      &1.7	      &1.4	      &0.9	      &1.6	      &1.1	      &1.4	      &1.0      \\
6364 [O~{\sc i}]	&1.8	      &1.2	      &0.6	      &0.4	      &0.8	      &0.5	      &0.3	      &0.2      \\
6548 [N~{\sc ii}]	&95.1	      &61.2	      &29.1	      &18.7	      &41.8	      &26.9	      &19.2	      &12.4     \\
6563 H$\alpha$		&339	      &217	      &439	      &281	      &383	      &246	      &442	      &284      \\
6583 [N~{\sc ii}]	&280	      &179	      &68.3	      &43.7	      &114	      &73.0	      &37.5	      &24.0     \\
6678 He~{\sc i}		&3.4	      &2.1	      &6.2	      &3.9	      &5.4	      &3.4	      &6.3	      &3.9      \\
6717 [S~{\sc ii}]	&33.7	      &20.9	      &10.4	      &6.5	      &15.3	      &9.5	      &7.0	      &4.4      \\
6731 [S~{\sc ii}]	&47.7	      &29.6	      &11.4	      &7.1	      &19.2	      &11.9	      &5.9	      &3.6      \\
7065 He~{\sc i}		&4.6	      &2.7	      &4.7	      &2.7	      &4.1	      &2.4	      &5.0	      &2.9      \\
7136 [Ar~{\sc iii}]	&25.2	      &14.4	      &23.3	      &13.3	      &21.4	      &12.2	      &23.4	      &13.4     \\
7237 [Ar~{\sc iv}]	&...          &...	      &1.2	      &0.7	      &...	      &...	      &0.9	      &0.5      \\
7281 He~{\sc i}		&...          &...	      &1.1	      &0.6	      &...	      &...	      &1.3	      &0.7      \\
7320 [O~{\sc ii}]	&22.5	      &12.4	      &4.0	      &2.2	      &7.6	      &4.2	      &1.5	      &0.8      \\
7330 [O~{\sc ii}]	&20.4	      &11.2	      &3.1	      &1.7	      &6.7	      &3.6	      &1.3	      &0.7      \\
7751 [Ar~{\sc iii}]	&6.6	      &3.3	      &5.9	      &3.0	      &5.8	      &2.9	      &6.2	      &3.1      \\
7774 O~{\sc i}		&...          &...	      &0.6	      &0.3	      &...	      &...	      &0.6	      &0.3      \\
8446 O~{\sc i}		&3.9	      &1.7	      &...	      &...	      &...	      &...	      &1.1	      &0.5      \\
8750 Pa12		&2.4	      &1.0	      &2.6	      &1.1	      &2.3	      &1.0	      &3.4	      &1.4      \\
8863 Pa11		&3.4	      &1.4	      &3.0	      &1.2	      &2.5	      &1.0	      &4.0	      &1.6      \\
9015 Pa10		&7.2	      &2.8	      &7.2	      &2.8	      &6.0	      &2.3	      &6.5	      &2.6      \\
9069 [S~{\sc iii}]	&68.1	      &26.6	      &66.3	      &25.9	      &59.4	      &23.2	      &86.9	      &34.0     \\
9229 Pa9		&6.9	      &2.6	      &7.2	      &2.7	      &5.8	      &2.2	      &7.6	      &2.9      \\
9532 [S~{\sc iii}]+Pa8	&219$^b$      &79.7$^b$       &215	      &78.5	      &145	      &52.8	      &300	      &110      \\
9711 [Fe~{\sc ii}]	&16.1	      &5.7	      &13.2	      &4.7	      &12.4	      &4.4	      &9.3	      &3.3      \\ \hline
\end{tabular}
\scriptsize 

$^{a}$ Observed line intensities were dereddened using the extinction
law of Cardelli et al.\ (1989), with $E(B-V)$=0.37 and $R_V=4.8$. \\
$^{b}$ The intensity of [S~{\sc iii}] $\lambda$9532 was assumed to be
3$\times$[S~{\sc iii}] $\lambda$9069 in the IF spectrum, since that
line at that position partly fell on a bad pixel on the detector.
\end{minipage}
\end{table*}

\subsection{Emission Line Morphology of the Finger}

The overall morphology of the Finger in Figures 2 and 3 is quite
striking, and consistent with the interpretation that the emission
arises from the ionized evaporating flow and dense ionization fronts
of an externally-illuminated molecular globule.  The [O~{\sc iii}],
H$\alpha$, and [S~{\sc ii}] images are reminiscent of many {\it
HST}/WFPC2 images of dust pillars and globules in other H~{\sc ii}
regions (e.g., Hester et al.\ 1996; Reipurth et al.\ 2003).
Differences in the observed structure in various emission lines give
interesting clues to the nature of the evaporating globule: In [O~{\sc
iii}] emission, the Finger is seen almost exclusively as a silhouette
against the bright background, except for a faint halo from the
evaporated flow around the globule.  In H$\alpha$, the emission from
the evaporating halo material and extinction features are almost
identical to [O~{\sc iii}], as the extended halo mostly disappears in
the H$\alpha\div$[O~{\sc iii}] image in Figure 4$a$.  H$\alpha$ is
significantly stronger than [O~{\sc iii}] only near the sharp
limb-brightened edges of the cloud, where strong [S~{\sc ii}] emission
is also seen at the ionization front (i.e. the H$\alpha\div$[O~{\sc
iii}] flux ratio image in Figure 4$a$ looks nearly identical to the
[S~{\sc ii}] image).  The globule itself appears to be optically
thick, especially at visual wavelengths where it totally obscures the
background H~{\sc ii} region; the emission from [O~{\sc iii}] and
H$\alpha$ seen projected within the boundaries of the globule is due
primarily to the ionized evaporating flow on the near side. The
globule does not become transparent at near-IR wavelengths like the
pillars in M16 (e.g., Thompson et al.\ 2002; McCaughrean \& Andersen
2002).  It is seen clearly as a silhouette in Pa$\beta$, He~{\sc i}
$\lambda$10830, H$\alpha$, and [O~{\sc iii}], but is not a silhouette
in [S~{\sc ii}] or H$_2$ because the emission from the globule's front
surface is brighter than the background.  This is especially clear
from the [S~{\sc ii}]$\div$H$\alpha$ ratio image in Figure 4$b$.  The
spatially-resolved structure of the stratified ionization fronts and
photoevaporating halo will be discussed in more detail in \S 5.

There appear to be two dominant locations of active photoevaporation
where the surface of the globule is directly irradiated by individual
stars; these are at the southern end of the globule (the ``knuckles'',
if the Finger globule is imagined to resemble a human hand), and at
the western side of the northern part (the ``wrist'').  The activity
in these two regions is highlighted best by emission from He~{\sc i}
$\lambda$10830 (Figure 3$a$) and H$_2$ (Figure 3$c$).  The He~{\sc i}
line has a metastable lower level and is enhanced by collisions in
dense ionized gas, such as that found in a dense ionized
photoevaporative flow.  H$_2$ emission will be brightest in the
photodissociation region (PDR) immediately behind the strongest
ionization fronts, where a strong flux of FUV (Balmer continuum)
photons penetrates, or where the gas is heated by a shock or
Ly$\alpha$ from the ionization front itself.  Both He~{\sc i} and
H$_2$ are most prominent in the two regions mentioned above: the
``knuckles'' and the ``wrist''.  Relatively weak limb-brightened
[S~{\sc ii}] emission is seen around the periphery of the globule,
probably caused by the ambient UV field and Ly$\alpha$ radiation.

The most obvious photoevaporating flow on the southern edge of the
globule exhibits a thin straight finger protruding normal to a bright
edge-on ionization front.  This thin finger appears to identify its
source of irradiation: It points toward P.A.=208$\arcdeg$, close to
the massive star WR25 (HD~93162) at P.A.=219$\arcdeg$ relative to the
Finger, although other early-type stars are seen along the same
apparent trajectory.  The ionizing source is addressed quantitatively
in \S 5.2.  Extending toward the south from this main ionization front
is a large-scale ionized flow, prominent in [O~{\sc iii}], H$\alpha$,
and Pa$\beta$, where it fills the lower left corner in Figures 2$a$,
2$b$, and 3$b$.  About 6$\arcsec$ from the end of the finger, one can
see an arc-shaped structure in these same emission lines; the density
here might be enhanced in a shock as the ram pressure of the
photoevaporating flow off the globule balances the pressure of the
ambient ionized gas in the H~{\sc ii} region.  From conservation of
momentum, this should occur at a distance from the globule's center
given roughly by $r_{if} (n_{if}/n_{H_{II}})^{0.5}$, where $n_{if}$
and $n_{H_{II}}$ are the electron densities at the ionization front
and in the H~{\sc ii} region, respectively, and $r_{if}$ is the radius
of curvature of the ionization front.  Electron densities are
estimated below (Table 3), and the resulting distance is in reasonable
agreement with the location of the shock.

Both of the main evaporating surfaces of the globule are found at
relatively flat interfaces (a large effective radius of curvature).
This may be an important factor in their enhanced emission, as the
radial density profile will fall off more rapidly in a spherically
divergent flow than from a planar ionzation front.  The northern
evaporative flow (the ``wrist'') is particularly interesting, as it
may be caused by a very nearby star -- the bright B star located only
10$\arcsec$ to the west (Tr16-207; see Figure
1).\footnotemark\footnotetext{Additionally, we note that the northern
part of the globule appears to be curved with a radius of curvature of
$\sim$10$\arcsec$, and with Tr16-207 located at the center of
curvature.}  The motivation for this conjecture is that both the
He~{\sc i} and H$_2$ emission are localized, while most of the rest of
the western side of the globule also appears relatively flat and
normal to the same direction.  If the ionizing source were a luminous
star at a large distance, we might expect to see a similar level of
enhanced emission all along the western side of the globule.  A third
region of active photoevaporation may have a different UV source from
the other two just mentioned; the southeast edge of the globule shows
enhanced emission in [S~{\sc ii}] and H$_2$, most notably on the east
side of the thin finger, while such emission is not as strong on the
west side of the same finger.  The very luminous and massive evolved
star $\eta$ Carinae is located toward the southeast; $\eta$ Car may
have been a prodigious FUV source in the recent past (more than 160
years ago) before it was surrounded by its young circumstellar dust
shell (Smith et al.\ 1998).  The O4 V((f+)) star HDE~303308 is seen
toward the east/southeast as well, and may add to the FUV flux
striking the side of the Finger.

\subsection{Proplyd Candidates?}

The new {\it HST}/WFPC2 images also reveal a small group of distinct
cometary clouds located immediately west of the Finger, the largest
member of which has approximate dimensions of
0$\farcs$5$\times$2$\arcsec$ (about 1100$\times$4500 AU).  This object
closely resembles the numerous ``proplyd candidates'' seen throughout
the Carina Nebula (Smith et al.\ 2003a).  This object, which we denote
104430.2-593953 (following the naming convention of Smith et al.\
2003a), is smaller than many of the other cometary objects in Carina,
but is still larger than the proplyds seen in Orion and is apparently
seen primarily in extinction and limb-brightened emission from
H$\alpha$ and [S~{\sc ii}].  The nature of these objects remains
somewhat ambiguous (are they analogues of Orion's proplyds, starless
cometary clouds, or something in between?), but this image hints that
an {\it HST} imaging survey of Carina would uncover many such objects
that are missed in ground-based images.

\begin{figure}\begin{center}
\epsfig{file=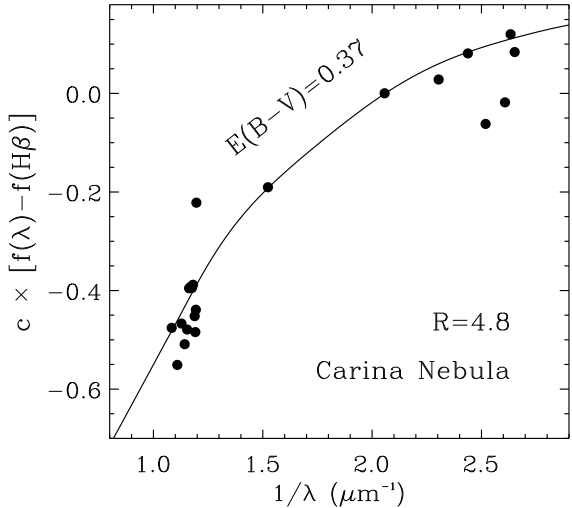,width=3in}\end{center}
\caption{Observed reddening for hydrogen lines in the background
Carina Nebula H~{\sc ii} region near the Finger, relative to H$\beta$
(dots).  The solid curve shows the reddening for $E(B-V)$=0.37 or a
logarithmic extinction at H$\beta$ of c$\simeq$0.83 (assuming the
ratio of total to selective extinction is $R_V$=4.8).}
\end{figure}

\section{SPECTROSCOPIC RESULTS}

Figure 5 shows ground-based spectra extracted from the long-slit data
for sub-apertures identified in Figure 1.  The IF spectrum does not
truly isolate emission from the thin ionization front on the surface
of the globule because these ground-based data lack the requisite
spatial resolution.  However, emission in this sub-aperture is
dominated by the limb-brightened emission from the IF; this is
especially true for low-ionization lines like [S~{\sc ii}], whereas
much of the emission in high-ionization lines like He~{\sc i} and
[O~{\sc iii}] may originate in immediately adjacent regions.  The IBL
spectrum samples the more extended photoevaporated flow, or the {\it
ionized boundary layer} that absorbs the hard UV photons incident upon
the globule.  The TOT spectrum includes both the IF and the IBL, and
is mainly useful for comparison with some diagnostic line ratios
mentioned later.  Finally, the background spectrum of the Carina
Nebula is shown for comparison; of course, this is a mixture of the
high-ionization gas inside the H~{\sc ii} region with the ionization
front and photodissociation regions on the far side of the nebula.

The most notable aspect of the IF spectrum is its low ionization, with
strong [O~{\sc ii}] $\lambda\lambda$3726,3729, [N~{\sc ii}]
$\lambda\lambda$6548,6583, and [S~{\sc ii}] $\lambda\lambda$6717,6731
compared to H$\alpha$.  These lines are typically enhanced at the
ionization front or just beyond it if they are not collisionally
de-excited; O has roughly the same ionization potential as H, the
ionization potential of S is below 13.6 eV, and although the
ionization potential of N is 14.5 eV, it can be ionized from the
excited $^2$D state by photons with energies as low as 12.1 eV.  The
IBL spectrum is intermediate in ionization and excitation between the
IF and the background H~{\sc ii} region, as expected.  The IF and IBL
spectra qualitatively resemble the low- and high-excitation spectra,
respectively, of the similar photoevaporating globule in the nearby
open cluster NGC~3572 (Smith et al.\ 2003b).

Table 2 lists observed and dereddened line intensities relative to
H$\beta$=100 for the various selected regions.  The reddening used to
correct the observed line intensities was determined by comparing
observed strengths of hydrogen lines in the background H~{\sc ii}
region to the Case B values calculated by Hummer \& Storey (1987).  As
shown in Figure 6, the observed Balmer and Paschen decrements suggest
a value for $E(B-V)$ of roughly 0.37$\pm$0.03, using the reddening law
of Cardelli, Clayton, \& Mathis (1989) with $R_V = A_V \div E(B-V) \
\approx \ 4.8$, which is appropriate for local extinction from dust
clouds around the Keyhole Nebula (Smith 1987; Smith 2002).  The value
of $E(B-V)$=0.37 we derive from nebular emission is close to the
average value of 0.47, and within the range of values of 0.25 to 0.64,
as listed for several bright O-type stars in Tr16 by Walborn (1995).
Table 3 lists representative physical quantities like electron density
and temperature derived from a standard deductive nebular analysis of
the usual line ratios.  These are useful to guide the interpretation
of the physics of the ionization front and photoevaporative flow
discussed below.

\begin{table}
\caption{Parameters derived from Dereddened Line Intensities}
\begin{tabular}{@{}llccc} \hline\hline
Parameter   &units    &IF     &IBL   &H{\sc ii} \\  \hline
n$_{\rm e}$ [S~{\sc ii}]   &cm$^{-3}$  &2200$\pm$500   &860$\pm$200   &230$\pm$70     \\
T$_{\rm e}$ [N~{\sc ii}]   &K          &9700$\pm$400   &9800$\pm$400  &12000$\pm$400  \\
T$_{\rm e}$ [O~{\sc iii}]  &K          &9700$\pm$700   &10000$\pm$700 &15200$\pm$1300 \\
T$_{\rm e}$ [S~{\sc iii}]  &K          &10000$\pm$1500 &8000$\pm$1100 &7200$\pm$900   \\  \hline
\end{tabular}
Values for TOT from lines in Table 2 are not shown here, because TOT
contains distinct regions (IF and IBL) with different physical
parameters.
\end{table}

\section{ANALYSIS OF THE PHOTOEVAPORATIVE FLOW}

Combining spectra with WFPC2 images that spatially resolve the
stratified IF on the globule's surface provides a powerful diagnostic
tool, and allows us to undertake a detailed analysis of the
photoevaporative flow and the incident UV field.  From our narrowband
images we made tracings of the observed intensity normal to the IF at
five different positions around the Finger, indicated as $a$ through
$e$ in Figure 7.  The corresponding intensity tracings at these five
postions are shown in Figure 8$a$ to $e$.  Surface brightness in
Figure 8 has been corrected for reddening and extinction, with
$E(B-V)$=0.37 and $R$=4.8.

\begin{figure}\begin{center}
\epsfig{file=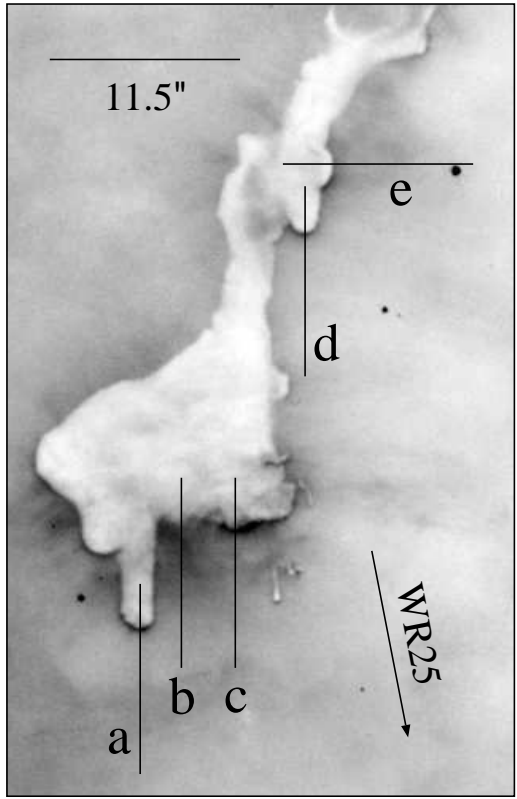,width=2.7in}\end{center}
\caption{Locations of spatial intensity tracings superposed on the
{\it HST}/WFPC2 F656N image of the Finger, rotated so that
P.A.=208$\arcdeg$ (the P.A. of the thin finger) is vertical.  The
positions labeled $a$ through $e$ correspond to tracings in Figure 8.}
\end{figure}

\subsection{The Stratified Ionization Fronts of the Finger}

All the tracings in Figure 8 show the same general stratification with
respect to the IF (at position=0), although there are some detailed
differences.  Starting from the left, H$_2$ emission peaks 0$\farcs$5
to 1$\farcs$5 (1200 to 3500 AU at d=2.3 kpc) behind the IF, while
[S~{\sc ii}], H$\alpha$, and Pa$\beta$ then rise sharply.  [S~{\sc
ii}] peaks at the IF and decreases exponentially thereafter, while
hydrogen lines continue to rise, with their peak $\la$0$\farcs$25
outside the IF in most cases (except position $b$).  [O~{\sc iii}] and
He~{\sc i} rise sharply at the IF (due to extinction by dust), reach a
broad peak (if any) outside the IF, and then decrease gradually at
larger distances.  The narrowest [S~{\sc ii}] zones with a FWHM of
$\sim$0$\farcs$5 are spatially resolved by {\it HST}, but the widths
of some of the narrow Pa$\beta$ and He~{\sc i} zones are unresolved in
the ground-based images.

Far from the IF ($\sim$7$\arcsec$) the intensities of H$\alpha$ and
[O~{\sc iii}] are roughly equal in each set of tracings.  Then, moving
toward the IF, the intensities of H$\alpha$ and [O~{\sc iii}] diverge
as hard photons capable of ionizing O$^+$ to O$^{2+}$ (with $h\nu>$35
eV) are eaten up by the photoevaporating flow and O$^{2+}$ recombines
(note the very strong [O~{\sc ii}] $\lambda\lambda$3727,3729 at the IF
in Figure 5 and Table 2).  At the same time, the strength of [S~{\sc
ii}] tends to increase in proportion to the difference between
H$\alpha$ and [O~{\sc iii}].  This is consistent with the qualitative
similarity between the [S~{\sc ii}] image (Figure 2$c$) and the
[O~{\sc iii}]$\div$H$\alpha$ ratio (Figure 4$a$).  Although every
position shares the same basic behavior described above, there are
differences among the five positions.  In particular, the IF's of $a$
and $d$ are much narrower than the other positions.  The
stratification at position $b$ is particularly broad.  These
differences are probably due to different radii of curvature (although
position B may be affected by evaporation off the near side of the
thin finger as well).  For example, at $b$ the IF appears nearly
straight, in stark contrast to the small radius of curvature at the
end of the finger at position $a$.  In general, a smaller radius of
curvature means that the photoevaporative flow diverges and the
density drops faster than in a planar IF, so the stratified ionization
structure is compressed.  Both positions $b$ and $e$ seem to be nearly
planar surfaces with a high-density flow extending far from the IF.

\begin{figure*}\begin{center}
\epsfig{file=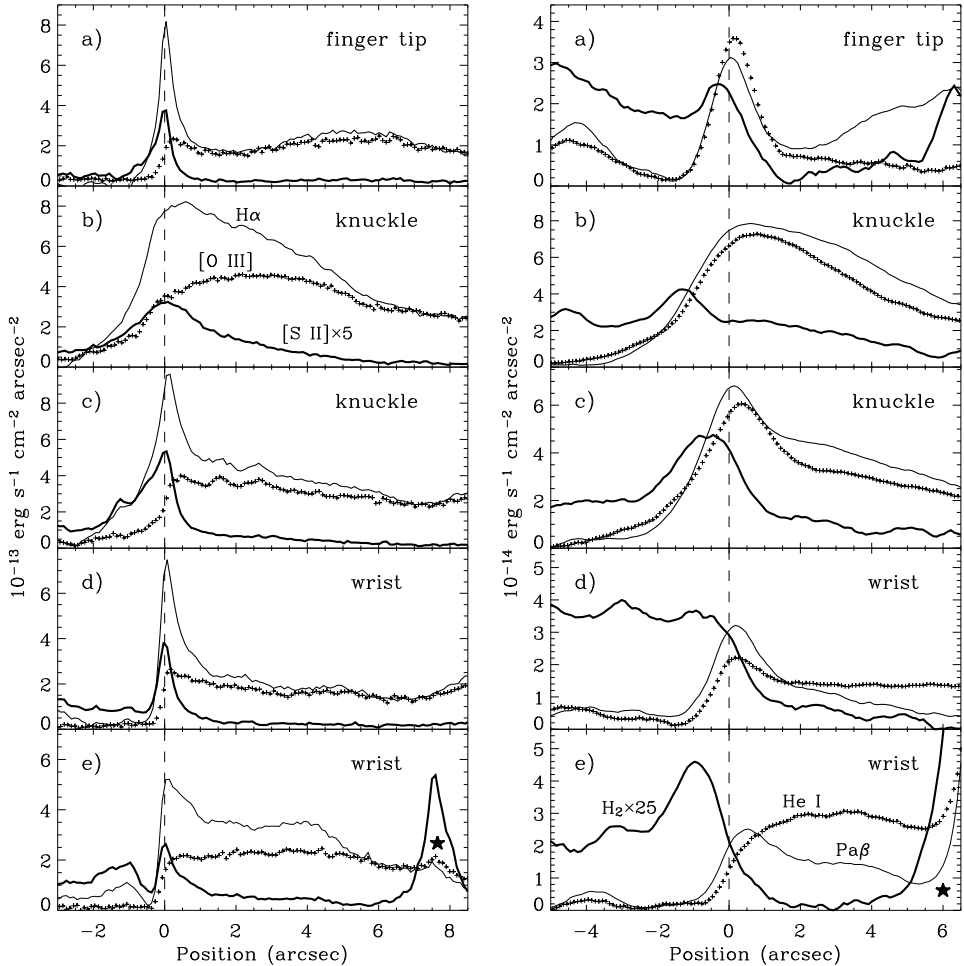,width=5.1in}\end{center}
\caption{Intensity tracings of the ionization fronts and
photoevaporative flows from the surface of the globule.  Panels $a$
through $e$ correspond to the positions shown in Figure 7.  The left
side shows intensity tracings made from {\it HST}/WFPC2 images in the
[O~{\sc iii}] $\lambda$5007 (F502N; crosses), H$\alpha$ (F656N; thin
line), and [S~{\sc ii}] $\lambda\lambda$6717+6731 (F673N; thick line)
filters.  The right shows tracings made from ground-based near-IR
images in the light of He~{\sc i} $\lambda$10830 (crosses), hydrogen
Pa$\beta$ (thin line), and H$_2$ $v=1-0$ S(1) $\lambda$21218 (thick
line).  All fluxes have been corrected for extinction and reddening
with $E(B-V)$=0.37 and $R$=4.8, as shown in Figure 6.  In each filter,
a constant value was subtracted corresponding to the emission from the
background H~{\sc ii} region far from the globule.  Note that the
weaker lines of [S~{\sc ii}] and H$_2$ have been multiplied by factors
of 5 and 25, respectively, for more effective comparison.  The
horizontal axis shows spatial position relative to the peak of the
[S~{\sc ii}] emission, adopted as the nominal location of the
ionization front.  The star symbol in panel (e) marks the position of
a star (see Figure 7).}
\end{figure*}

\begin{table}
\caption{Ionization Front Properties}
\begin{tabular}{@{}lccccl}\hline\hline
Pos. &$S_{H\alpha}$ &$L$ &$n_e$ &log$_{10}$($\Phi$) &Spectral \\
 &($^a$) &(10$^{16}$ cm) &(cm$^{-3}$) &(s$^{-1}$ cm$^{-2}$) &Type$^b$ \\ \hline
a...  &6.5  &3.5   &6200  &11.06  &O7.5  \\
b...  &5.5  &10.4  &3300  &10.99  &O6.5  \\
c...  &7.0  &6.9   &4600  &11.09  &O7  \\
d...  &6.0  &3.5   &6000  &11.03  &O7.5  \\
e...  &3.0  &10.4  &2400  &10.73  &$\la$B0  \\ \hline
\end{tabular}
$^a$ 10$^{-13}$ erg s$^{-1}$ cm$^{-2}$ arcsec$^{-2}$. This H$\alpha$
intensity may be contaminated somewhat by [N~{\sc ii}] $\lambda$6583
incuded in the F656N filter; this effect is strongest at the IF.
Using the [N~{\sc ii}]/H$\alpha$ ratio for the IF spectrum from Table
2, combined with the transmission curve of the F656N filter of {\it
HST}/WFPC2, we estimate that [N~{\sc ii}] contributes less than about
5\% of the measured H$\alpha$ flux at the IF. \\ $^b$ See Sankrit \&
Hester (2000).
\end{table}

\subsection{Does WR~25 Ionize the Finger?}

As noted earlier, the finger at the southern end of the globule seems
to point toward the luminous WNL star WR25 (HD~93162), but since some
other luminous early-type stars are seen along this same general
trajectory, this relationship should be tested quantitatively.  By
examining the properties of the IF and photoevaporative flow from the
globule, we can estimate the necessary ionizing photon flux and the
shape of the incident continuum.  These can then be compared to likely
values for candidate sources.

The electron density at the IF is the critical quantity for estimating
the flux of ionizing photons striking the globule if we assume that
the recombination rate roughly balances ionizations.  From the ratio
[S~{\sc ii}] $\lambda$6717$\div\lambda$6731 we find $n_e \simeq$ 2200
cm$^{-3}$ in the thin [S~{\sc ii}] emitting region on the surface of
the globule (see Table 3).  However, this is only the value near
position $b$, which may be somewhat unusual because of its nearly
planar geometry, as noted earlier, and ground-based spatial resolution
in these spectra does not effectively sample the highest density at
the thin IF.  At $b$ and other positions $a$, $c$, $d$, and $e$, the
electron density at the IF can be estimated independently using the
extinction-corrected H$\alpha$ surface brightness $S_{H\alpha}$ in
Figure 8.  If we assume $n_e \simeq n_p$, and adopt a reasonable
estimate of the depth of the emitting
layer\footnotemark\footnotetext{The values of $L$ are determined from
images in a somewhat subjective manner, and the uncertainties will be
smallest for the ionization fronts at positions $a$ and $d$, which
have the least ambiguous geometry.} along the line of sight $L$, then
we have

\begin{equation}
n_e^2 = \frac{4 \pi (S_{H\alpha}\times206265^2) }{\alpha_{H\alpha}^{\rm eff} h\nu_{H\alpha} L}
\end{equation}

\noindent where $\alpha_{H\alpha}^{\rm eff} = 8.6\times$10$^{-14}$
cm$^3$ s$^{-1}$ for T$\simeq$10$^4$ K (Osterbrock 1989), and
206265$^2$ just converts from arcsec$^{-2}$ to sr$^{-2}$ so that we
can simply use values gleaned from Figure 8.  Adopted values of
$S_{H\alpha}$, $L$, and $n_e$ are listed in Table 4.  Since $n_e$
derived from images is somewhat higher than that derived from the
[S~{\sc ii}] lines for position $b$, as we would expect if [S~{\sc
ii}] lines trace a region where hydrogen already has a large neutral
fraction, we take the values for $n_e$ in Table 4 as more reliable
estimates of the IF density.  We can then use these values of $n_e$ to
estimate the number of ionizing photons $\Phi$ incident on an IF of
radius $r$ per second and per unit area if we assume that ionizations
are approximately balanced by recombinations at the IF, so that we
have

\begin{equation}
\Phi \ \simeq \ \Delta r \ \alpha_B \ n_e^2
\end{equation}

\noindent where $\alpha_{B} = 2.59\times$10$^{-13}$ cm$^3$ s$^{-1}$ is
the total case B hydrogen recombination coefficient for
T$\simeq$10$^4$ K (Osterbrock 1989).  To calculate $\Phi$ for the five
positions listed in Table 4, we assumed the characteristic thickness
of the recombination front is $\Delta r \simeq L \div 3$, which seemed
a fair approximation based on the width of the limb-brightened
surfaces seen in images.  Judging by the morphology observed in Figure
2, positions $a$ through $d$ all have the same ionizing source, and
Table 4 indicates that all these positions have roughly the same
incident flux of ionizing photons, with an average of
log$_{10}$($\Phi$)=11.04 s$^{-1}$ cm$^{-2}$.  Position $e$ has a lower
incident UV flux, with log$_{10}$($\Phi$)$\simeq$10.7 s$^{-1}$
cm$^{-2}$.  The disagreement between the values of $\Phi$ for
positions $a$ through $d$ compared to $e$ is consistent with the
apparent morphology in Figure 2, where position $e$ seems to have a
different UV source.

The shape and relative intensities of various lines in Figure 8 can
also be used to constrain the spectral type of the ionizing star, by
comparison to detailed emission models of photoevaporative flows.
Sankrit \& Hester (2000; SH hereafter) modeled ionized
photoevaporative flows as seen in the same optical emission line
tracers as we observed with {\it HST}.  SH found that while the
maximum density at the IF depends on the total ionizing flux but not
on the shape of the incident continuum, the ratio of [O~{\sc iii}] to
H$\alpha$ is sensitive to the spectral type of the ionizing star.  In
Table 4 we list spectral types of the ionizing stars inferred from
values of the peak [O~{\sc iii}]$\div$H$\alpha$ ratio for positions
$a$ through $e$, corresponding to the models of SH with
log$_{10}$($\Phi$)$\simeq$11 s$^{-1}$ cm$^{-2}$ for positions $a$
through $d$, and log$_{10}$($\Phi$)$\simeq$10.7 s$^{-1}$ cm$^{-2}$
for position $e$ (see their Figure 8).  We regard these characteristic
spectral types as very rough estimates; they should be taken as
representing the approximate ratio of photons with $h\nu\ga$35 eV to
those with $h\nu\ga$13.6 eV, rather than the actual spectral type of a
main sequence star (SH also noted that the derived properties differed
depending on the type of model atmosphere
used).\footnotemark\footnotetext{When interpreting the characteristic
spectral types, one should also keep in mind recent revisions to the
$T_{eff}$ scale for O stars (e.g., Martins et al.\ 2002).}  This is
especially relevant to the possibility of WR25 as the ionizing source,
as it is not a main-sequence star and has a strong wind
($\dot{M}=10^{-4.4}$ M$_{\odot}$ yr$^{-1}$; Crowther et al.\ 1995).
Also, these characteristic spectral types may underestimate the actual
hardness of the source energy distribution, since hard photons may be
absorbed in the ambient material between the star and the Finger or by
dust in the photoevaporative flow itself.  In general, the peak
[O~{\sc iii}]$\div$H$\alpha$ ratios for positions $a$ through $d$ are
consistent with a characteristic spectral type of $\sim$O7 or O7.5,
according to the models by SH, while position $e$ requires a source
with a spectral type later than B0 (an [O~{\sc iii}] peak is weak or
non-existent at position $e$).

\begin{table}
\caption{Candidate UV Sources}
\begin{tabular}{@{}lcccc}\hline\hline
Star &Spectral &P.A.  &$R$      &log$_{10}$(Q$_H$)  \\
     &Type     &(deg) &(parsec) &(s$^{-1}$) \\ \hline
WR25                 &WNL          &219   &2.82  &50.1    \\
Tr16-244             &O4 If        &215   &2.71  &49.9    \\
HD 93205             &O3.5 V ((f)) &176   &2.96  &49.5    \\
HD 93204             &O5 V ((f))   &179   &3.12  &49.2    \\
-59$\arcdeg$2574     &B1-B1.5 V    &209   &0.87  &46.5    \\ \hline
\end{tabular}

Spectral types are taken from Walborn et al.\ (2002) and Massey \&
Johnson (1993), $R$ is the projected separation from the Finger
measured in images, and ionizing photon fluxes for spectral types are
taken from Smith et al.\ (2002), except for WR25 where Q$_H$ is from
models by Crowther et al.\ (1995).  Also, HD~93205 is a binary,
actually classified as O3.5 V ((f)) + O8 V, but the hotter component
dominates the ionizing luminosity.
\end{table}

Figure 9 shows the hydrogen-ionizing photon luminosity Q$_H = 4 \pi
R^2 \Phi$ that would be required at a distance $R$ to produce a flux
of log$_{10}$($\Phi$)=11.04 s$^{-1}$ cm$^{-2}$ at the surface of the
Finger.  This is compared to the actual values of Q$_H$ and $R$ for
stars that are potential sources of the UV flux evaporating the
Finger.  $Q_H$ values are taken from Smith et al.\ (2002) for the
corresponding spectral type.  For WR25 we adopted $Q_H$ from the
models by Crowther et al.\ (1995; Crowther, private comm.; models by
Smith et al.\ 2002 and Crowther et al.\ 1995 include line blanketing.)
The positions of these candidate source stars are identified in Figure
1 and their properties are collected in Table 5.  Figure 9 shows two
values for $R$: the separation on the sky given in Table 5, and a
value that is a factor of $\sqrt{2}$ larger to account for a possible
projection effect.  From Figure 9 we conclude that the dominant source
of ionization for the Finger is either the WNL star WR25 (Crowther et
al.\ 1995) or the neighboring O4~If star Tr16-244 (star number 257 of
Massey \& Johnson 1993) --- or perhaps both working together.  It is
somewhat unsettling that the thin Finger appears to point toward a
slightly skewed position angle of $\sim$208$\arcdeg$, while WR25 is
found at P.A.=219$\arcdeg$ and Tr16-244 is at P.A.=215$\arcdeg$,
relative to the Finger; one might expect a feature like the Finger to
point {\it directly} toward its UV source.  However, other factors may
affect the appearance of photoionized columns as well (see Williams et
al.\ 2001; Cant\'{o} et al.\ 1998).  The Finger does point almost
exactly at CPD $-$59$\arcdeg$2574, but as an early B star, it does not
even come close to providing the required ionizing flux (Figure 9).
Located directly south of the Finger, the massive hot stars HD~93204
and HD~93205 also provide insufficient ionization.

\begin{figure}\begin{center}
\epsfig{file=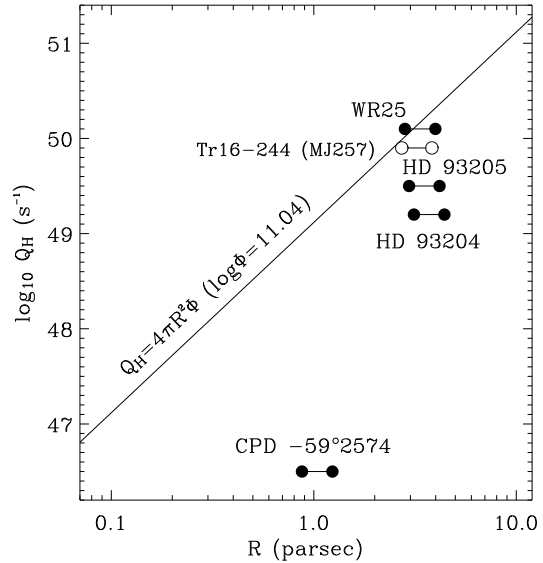,width=3.2in}\end{center}
\caption{The solid line labeled $Q_H = 4 \pi R^2 \Phi$ shows the
number of hydrogen-ionizing photons Q$_H$ that a star must emit at a
separation R$_{\rm pc}$ in order to ionize the Finger.  Q$_H$ is
plotted for several stars including WR25 at two different distances;
the near distance is just the projected separation on the sky, and the
far distance is an arbitrary factor of $\sqrt{2}$ larger.}
\end{figure}

One further complication, however, is that the ionizing continuum
striking the Finger appears to be softer than that produced by either
WR25 or Tr16-244.  The spectral type inferred from the [O~{\sc
iii}]$\div$H$\alpha$ ratio at the IF is only about O7 (Table 4),
whereas both likely UV sources have much earlier spectral types.  We
have not considered the effect of the strong X-ray flux from WR25
(e.g., Raassen et al.\ 2003; Seward et al.\ 1979).  Since there are no
other plausible sources of ionization in this direction, this suggests
that the incident radiation field is softened as it passes through the
nebula and the IBL on its way to the Finger.  From the spectrum of the
IBL we would derive an earlier spectral type from the [O~{\sc
iii}]$\div$H$\alpha$ ratio than in the thin IF itself (Figure 5 and
Table 2).

\subsection{Mass-Loss Rate and Lifetime}

From $^{12}$CO(2-1) observations, Cox \& Bronfman (1995) estimated a
characteristic mass for the Finger of $\sim$6 M$_{\odot}$, but because
the emission sampled mostly the warm surface of the cloud, they noted
that the total mass of the globule is probably 10 to 20 M$_{\odot}$.
This mass provides the reservoir for future photoevaporation, and
allows us to estimate a likely evaporation timescale for the cloud.
If the evaporation of the cloud is dominated by the brightest southern
IF, and if the geometry can be approximated as a cylinder illuminated
on one side, the mass loss rate is given roughly by

\begin{equation}
\dot{M} \simeq \pi r^2 m_H n_H v 
\end{equation}

\noindent where $r$ is the characteristic radius of curvature of the
cloud appropriate for the apparent size of the IF ($\sim$10$^{17.3}$
cm), and $v$ is the speed of the evaporative flow through the IF.  The
electron density at the main evaporating surface (positions $c$ and
$d$) is $\sim$4000 cm$^{-3}$ (Table 4), while right at the IF itself
we typically have $n_{\rm e} \simeq 0.7 n_{\rm H}$ because the gas is
not fully ionized (e.g., SH).  If we assume that the material expands
away from the ionization front at the isothermal sound speed
($v\approx$10 km s$^{-1}$), then $\dot{M} \approx 2 \times 10^{-5}$
M$_{\odot}$ yr$^{-1}$.  In that case, the remaining lifetime of the
entire cloud is 10$^{5.3}$ to 10$^6$ years; about 15\% to 30\% of the
age of the nebula.  The Finger and other associated molecular clumps
probably represent the last vestiges of the original molecular cloud
core that spawned Tr16, while significant reservoirs for future star
formation exist in the neighboring giant molecular cloud (Grabelsky et
al.\ 1988).

\section{STAR FORMATION IN THE FINGER?}

In addition to the emission-line images described above, near-IR
images in the $J$, $H$, and $K$ \ broadband filters were obtained.  A
few reddened sources with $m_K\simeq$14 were detected within the
boundaries of the globule, but these were positioned randomly and
could easily be chance alignments of background sources (these point
sources can be seen in the 2.122~$\micron$ H$_2$ image in Figure
3$c$).  A few of these $K$-band sources seem preferentially located
near the limb-brightened edges of the globule, but no reddened stars
were detected near the southern ionization front or inside the thin
``finger'' protruding from it, as one might have expected.  However,
that doesn't {\it necessarily} mean that no stars are currently
forming here or that no stars will form in the future.  Below we
estimate the amount of extinction in the Finger and the $K$-band
source that could be hidden, as well as other pertinent quantities
like the column density and mass in the Finger and its associated
globule.  Then we discuss implications for current and future star
formation in this globule.

\begin{figure}\begin{center}
\epsfig{file=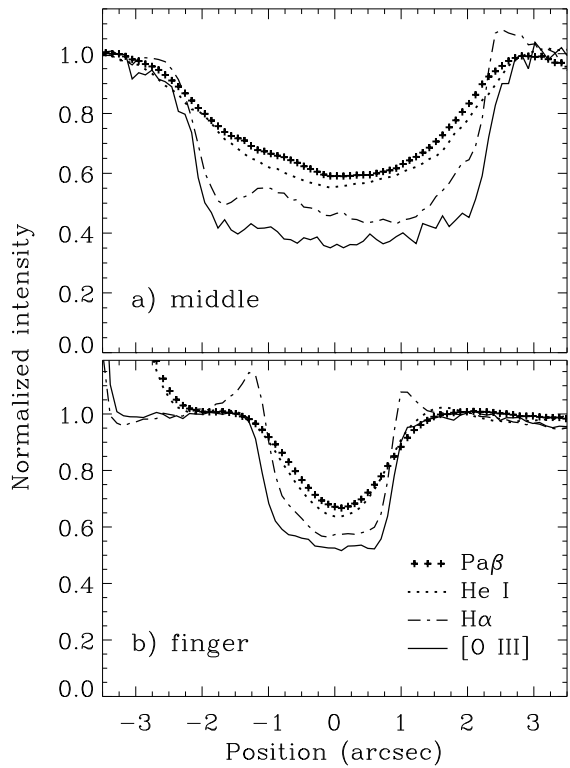,width=3in}\end{center}
\caption{Normalized intensity cuts (a) through a narrow section near
the middle of the globule and (b) through the thin extended finger for
several emission lines.  These tracings are useful for assessing the
extinction in the globule (see text \S 6.1).}
\end{figure}

\subsection{Extinction and Dust Mass}

The Finger globule is seen as a silhouette in [O~{\sc iii}],
H$\alpha$, He~{\sc i} $\lambda$10830, and Pa$\beta$, so tracings
through the globule in these emission lines can be used to estimate
the optical depth, mass column density, and total mass of the globule.
Figure 10 shows representative cross-cuts through the middle of the
globule and through the thin finger (perpendicular to tracing ``a'' in
Figure 7).  Differences in spatial resolution between the {\it HST}
and ground-based data aside, it is clear that the minimum emission
intensity in the center of each tracing $I_{\lambda}$ increases with
wavelength.  Thus, the globule becomes translucent at the longer
near-IR wavelengths.  Interpreting these tracings is not entirely
straightforward, however, because the Finger is not a simple
silhouette blocking light from a background source --- instead, it is
a bright-rimmed globule suspended inside the H~{\sc ii} region.  Thus,
some fraction of the observed minimum intensity originates along our
line-of-sight to the Finger or near its surface.  For example, it
would not be unreasonable to assume that roughly half the intensity
for each line in Figure 10 is foreground emission.  Since the globule
has a flat-bottom profile and negligible limb brightening in [O~{\sc
iii}], we can make a crude correction for this foreground emission if
we assume that the globule is optically thick and ``black'' in the
[O~{\sc iii}] line.  Then, the effective transmission at a given
wavelength is given by $(I_{\lambda}-I_{O3})/(1-I_{O3})$, where
$I_{O3}$ is the minimum flux in the [O~{\sc iii}] line in the middle
of the globule.  Contamination from the IBL is probably worst in
H$\alpha$ and Pa$\beta$, so using the He~{\sc i} $\lambda$10830 line,
we estimate an effective transmission through the middle of the
globule of roughly 30\%, and a corresponding optical depth ($\tau$) of
$\sim$1.2 at a wavelength of 1.08 $\micron$.  Similarly, for the thin
finger we estimate a slightly higher $\tau$ of $\sim$1.5 at 1.08
$\micron$.  The dust-mass column density is given by

\begin{equation}
m_{_{DUST}} \simeq \frac{4 a \rho \tau}{3 \ Q_{abs}} 
\end{equation}

\noindent where $a \approx 0.1 \ \micron$ is the assumed grain radius
(large grains dominate the opacity), $\rho \approx$ 1 g cm$^{-3}$ is
the average grain mass density, and $Q_{abs}$=0.04 is the extinction
coefficient at 1.08 $\micron$ (e.g., Draine \& Lee 1984).  Then, with
$1.2 \la \tau \la 1.5$ at 1.08 $\micron$, we have characteristic
dust-mass column densities of 4 to 5$\times$10$^{-4}$ g cm$^{-2}$.  If
we integrate over the entire projected area of the globule, and assume
a typical gas:dust mass ratio of 100:1 (this is the dominant source of
uncertainty), then we find a total mass for the Finger globule of 6 to
8 M$_{\odot}$.  Likewise, the mass of the small clump at the end of
the thin finger is of order 0.1 to 0.2 M$_{\odot}$.  The corresponding
hydrogen column density through the globule is $N_H \approx$ 2 to
3$\times$10$^{22}$ cm$^{-2}$.

The excellent agreement with the mass of 6 M$_{\odot}$ derived for the
globule by Cox \& Bronfman (1995) is fortuitous, given the large
inherent uncertainty in the assumed gas:dust mass ratio.  In fact,
just as Cox \& Bronfman noted that their value was an underestimate
because the CO emission was dominated by warm outer layers of the
globule, ours is probably an underestimate because some parts of the
finger may be optically thick even at 1 to 2 $\micron$ and may hide
additional mass.

In our images, the $K$-band detection limit for point sources
projected against the uneven background emission associated with the
Finger globule is $m_K\simeq$17.  The extinction estimated above could
then hide an intrinsic $m_K\simeq$15.5 source from detection.  At 2.3
kpc, this corresponds to an absolute $K$ magnitude of roughly 3.5 to
4, or a young star of roughly 0.3 to 0.5 M$_{\odot}$.  (Note that this
is comparable to the minimum mass of dust and gas within the end of
the thin finger.)  Considering our detection limit we cannot rule out
the possibility that a low-mass star has already formed at the end of
the finger, so deeper $JHK$ images with high spatial resolution are
desirable.

\subsection{Future Star Formation in the Finger?}

Even though we have found no direct evidence for newly-formed stars
within the Finger, this globule and others like it in Carina may still
be likely sites of future star formation.  In \S 5.3 we found that the
remaining lifetime before the globule evaporates away is of order
10$^{5.3}$ to 10$^6$ years, which is longer than the average free-fall
time for the whole globule, and is plenty of time to allow a dense
clump to collapse and form a protostar that would survive exposure to
the interior of the H~{\sc ii} region.  Thus, it is worth a closer
look at the physical conditions within the globule and its advancing
IF.  In particular, we ask whether the formation of a star may be
triggered by radiation-driven implosion.  We consider two cases: the
main southern IF working on the body of the globule, and the end of
the thin finger protruding from it.

\subsubsection{The Main Ionization Front}

With a characteristic radius of $\sim$10$\arcsec$ or
$\sim$3$\times$10$^{17}$~cm, the Jeans mass for the whole globule is
$\sim$2 M$_{\odot}$ (T/10 K).  Thus, with a temperature of 10 - 40 K
(e.g. Cox \& Bronfman 1995), a mass larger than 6 M$_{\odot}$ makes
it plausible that the globule can collapse to form stars --- or at
least, one could argue that the globule is on the verge of collapse
under self gravity, neglecting rotation and magnetic support.  Thus,
external pressure may play an important role in the globule's future.

When evaluating if radiation-driven implosion is important, we must
consider pressure balance between the ionization front (P$_{if}$) and
the pressure inside the neutral globule (P$_0$).  We have

\begin{equation}
P_{if} = n_e k T_e + m_H n_e v^2
\end{equation}

\begin{equation}
P_0 = \frac{\rho_0}{\mu m_H} k T + \rho_0 \sigma^2 + \frac{B^2}{8 \pi}
\end{equation}

\noindent where $m_H n_e v^2$ is the back pressure from the
photoevaporative flow launched from the IF (roughly equal to $n_e k
T$), $\rho_0$ is the mass density, $\mu$ is the mean molecular weight,
$\rho_0 \sigma^2$ is the inferred turbulent pressure of the molecular
gas, and the last term in equation (6) is the magnetic pressure inside
the cloud.  The electron density and temperature of the IF are roughly
4000 cm$^{-3}$ and 10$^4$ K (Tables 3 and 4), respectively, the
average mass density $\rho_0$ inside the cloud is
$\sim$2$\times$10$^{-19}$ g cm$^{-3}$ (see \S 6.1), and $\sigma$ is
roughly 1.1 km s$^{-1}$ inferred from CO line widths of 2.7 km
s$^{-1}$ (Cox \& Bronfman 1995).  If we neglect magnetic
fields\footnotemark\footnotetext{Note that turbulence dominates over
thermal pressure, and we have neglected rotation because the straight
edges of the thin finger protruding from the globule suggest that it
has maintained the same orientation for its evaporation timescale of
10$^5$ years.}, we find $P_{if}/P_0 \simeq 5$.  Thus, a significant
external overpressure may be causing the globue to collapse.  If
magnetic pressure is to support the globule, the perpendicular
component of the magnetic field that would be required is
$\sim$5$\times$10$^{-4}$ G.  Interestingly, in their analysis of the
pillars of the Eagle Nebula, White et al.\ (1999) find a similar field
of $B\simeq5.4\times10^{-4}$ G (in fact, many of the physical
parameters of the Finger resemble those estimated for the M16
pillars).  However, as noted by those authors, a field of this
strength would broaden the observed line width to $\ga$4 km s$^{-1}$
because of Alfvenic motions, which is broader than the CO linewidth of
2.7 km s$^{-1}$ reported by Cox \& Bronfman (1995) for the Finger.
Thus, we find it likely that radiaton-driven implosion at the IF may
be triggering the formation of low-mass stars in the globule.

\subsubsection{The Thin Finger}

The thin finger protruding from the IF presents a somewhat different
case, because it is smaller and denser than the main globule.  One
might expect that it was much denser than its intial surroundings, and
that it has come into pressure equilibrium with the IF, since it has
obviously resisted evaporation more than adjacent areas of the initial
cloud (see also Williams et al.\ 2001).  We approximate the end of the
thin finger as a spherical globule with radius 5$\times$10$^{16}$ cm,
a mass of $\ga$0.2 M$_{\odot}$ (see \S 6.1), and a corresponding mass
density of $\sim$10$^{-18}$ g cm$^{-3}$.  The Jeans mass is then
$\sim$0.3 M$_{\odot}$ (T/10 K), and again, we find it plausible that
the dense clump may be in the process of collapsing to form a star ---
especially if 0.2 M$_{\odot}$ is an underestimate of the clump's mass
because of high optical depth as noted earlier.  Following the
analysis above, we find $P_{if}/P_0 \simeq 1.4$.  This is considerably
more uncertain than for the main globule above, because turbulence
dominates the internal pressure, and we do not know to what extent the
value of $\sigma$ used above applies to the small fingertip (this
uncertainty also precludes a meaningful estimate of the magnetic field
needed to halt collapse).  Thus, the most likely scenario may be that
the fingertip is in pressure equilibrium, and that a slowly-advancing
D-type (dense) IF is eating into the fingertip, and the shock from the
IF has long ago passed through the clump.  If a star is to form here,
the process should already be underway.  The fingertip may contain a
faint low-mass protostar that has escaped detection in our images, or
it may have a very young Class~0 protostar.  In the latter case,
thermal IR emission from a dense hot core would be difficult to
detect, as it would be masked by thermal emission from the adjacent
warm IF and PDR.

\section{CONCLUSIONS}

We have presented optical narrow-band {\it HST}/WFPC2 images,
ground-based optical spectra, and near-IR images of the ``Finger'' --
a photoevaporating molecular globule in the core of the Carina Nebula.
The main conclusions of this work are the following:

1.  The Finger globule exhibits an interesting morphology, with a thin
    extended middle finger apparently pointing toward its source of
    ionizing photons.  The spatially-resolved structure of the
    stratified ionization fronts are consistent with the
    interpretation of the Finger as an optically-thick
    photoevaporating molecular globule, similar to structures often
    seen in {\it HST} images of H~{\sc ii} regions.

2.  Quantitatively, electron densities and the corresponding flux of
    ionizing photons incident upon the southward-facing ionization
    fronts of the Finger indicate that the dominant UV source is
    either WR25 (a late-type Wolf-Rayet star), Tr16-244 (O4 If), or
    perhaps both.  This is reassuring, since the finger points to
    within a few degrees of these stars.

3.  The mass-loss rate for the main evaporating surface of the globule
    is of order 2$\times$10$^{-5}$ M$_{\odot}$ yr$^{-1}$.

4.  From extinction measurements, we estimate an average hydrogen
    column density of a few times 10$^{22}$ cm$^{-2}$ through the
    globule, and a total mass (assuming a gas:dust mass ratio of
    100:1) of at least 6 M$_{\odot}$, in agreement with independent
    estimates from molecular studies.  This is an underestimate if the
    globule contains clumps that are optically-thick in the near-IR.

5.  The remaining lifetime of the globule before it is evaporated away
    is of order 10$^{5.3}$ to 10$^6$ years.

6.  Several reddened stars are seen projected within the boundaries of
    the Finger globule, but whether or not these sources are
    newly-formed stars that are physically associated with the globule
    is uncertain.  No reddened star is seen at the apex of the thin
    protruding finger or immediately behind the main ionization front,
    down to a limit of $m_K \simeq 17$.

7.  Considering the properties of the advancing ionization front, it
    appears likely that stars are currently forming or will soon form
    in the globule, triggered by radiation-driven implosion.  At the
    main ionization front we find an external overpressure of a factor
    of $\sim$5, and a smaller overpressure ($\ga$1) at the end of the
    thin finger.

\smallskip\smallskip\smallskip\smallskip
\noindent {\bf ACKNOWLEDGMENTS}
\smallskip
\scriptsize

We thank John Bally for supplying the ground-based [S~{\sc ii}] image
used in Figure 1, and we benefitted from helpful discussions with Paul
Crowther regarding models of WR25.  Support was provided by NASA
through grant HF-01166.01A from the Space Telescope Science Institute,
which is operated by the Association of Universities for Research in
Astronomy, Inc., under NASA contract NAS~5-26555.  NOAO funded N.S.'s
travel to Chile and accommodations while at CTIO.  Some travel support
was also provided by NASA grant NAG-12279 to the University of
Colorado.

\end{document}